# Highly Sensitive Gas Sensors Based on Silicene Nanoribbons

S. M. Aghaei [a], M. M. Monshi [a], and I. Calizo [a,b]

Inspired by the recent successes in the development of two-dimensional based gas sensors capable of single gas molecule detection, we investigate the adsorption of gas molecules ($N_2$, $NO$, $NO_2$, $NH_3$, $CO$, $CO_2$, $CH_4$, $SO_2$, and $H_2S$) on silicene nanoribbons (SiNRs) using density functional theory (DFT) and nonequilibrium Green's function (NEGF) methods. The most stable adsorption configurations, adsorption sites, adsorption energies, charge transfer, quantum conductance modulation, and electronic band structures of all studied gas molecules on SiNRs are studied. Our results indicate that $NO$, $NO_2$, and $SO_2$ are chemisorbed on SiNRs *via* strong covalent bonds, suggesting its potential application for disposable gas sensors. In addition, $CO$ and $NH_3$ are chemisorbed on SiNRs with moderate adsorption energy, alluding to its suitability as a highly sensitive gas sensor. The quantum conductance is detectably modulated by chemisorption of gas molecules which can be attributed to the charge transfer from the gas molecule to the SiNR. Other studied gases are physisorbed on SiNRs *via* van der Waals interactions. It is also found that the adsorption energies are enhanced by doping SiNRs with either B or N atom. Our results suggest that SiNRs show promise in gas molecule sensing applications.

## I Introduction

Although the global development of the chemical activities in the last century is a consequence of the human demands, it also affects human health and life quality from the associated environmental release of poisonous substances in the forms of solid, liquid and vapour. Therefore, gas detection is vital for managing chemical processes to prevent health hazards such as air pollution, device contamination, and for medical diagnosis. Today, developing novel gas sensing materials with high sensitivity, even down to the single molecule level, high selectivity, high stability, quick response and recovery, and low power consumption is of considerable importance. Graphene, the first discovered two-dimensional (2D) material,[1,2] renders outstanding properties such as high surface-volume ratio, low electronic temperature noise, remarkably high carrier mobility, high chemical and thermal stability, and fast response time, which make it promising in the development of ultrasensitive sensors with high packing density, higher sensitivity, better selectivity, faster recoverability, and less power consumption.[3-5] The potential application of graphene for gas sensors has been widely studied both experimentally[6,7] and theoretically.[8-10] However, the physisorption of common gas molecules such as $CO_2$, $CO$, $CH_4$, $H_2O$, $H_2$, $N_2$, $NO_2$, and $NO$ on pristine graphene[8] restrict its potential for single molecule detection.[11,12] It is found that sensing ability of graphene can be enhanced by introducing dopants or defects.[9,10,13]

Inspired by the alluring properties of graphene, silicene, the analogue of graphene for silicon and having a buckled honeycomb structure,[14-16] has garnered considerable interest because of its remarkable properties including ferromagnetism,[17] half-metallicity,[18] quantum Hall effect,[19] giant magnetoresistance,[20] and superconductivity.[21] Takeda and Shiraishi[14] reported silicene for the first time in 1994 and Guzmán-Verri and Voon[15] coined its name in 2007. Although free-standing silicene is not stable, it was experimentally fabricated on Ag,[21-24] Ir,[25] $ZrB_2$,[26] and $ZrC$.[27] Using first-principles calculations, various substrates such as h-BN,[28] SiC,[29] GaS,[30] graphene[31] and ZnS[32] which have weak van der Waals (vdW) interactions with silicene have been also studied to improve its stability. The extraordinary properties of silicene along with its compatibility with silicon based nanoelectronics may give the edge to silicene rather than graphene. Numerous potential applications of silicene in spintronics,[33] field-effect transistors (FETs),[34,35] and sensing devices[36,37] have been proposed. Unlike a flat graphene sheet, the silicene honeycomb structure is buckled[15,16] due to the tendency of silicon atoms to adopt $sp^3$ and $sp^2$ hybridization rather than only $sp^2$ hybridization.[22] This buckled structure makes it possible for the band gap of silicene to be tuned more intensively with an external electric field[38] and with binding adsorbates[39] as compare to graphene.[40] Furthermore, silicene shows a considerably higher chemical reactivity for atoms[41-46] and molecules[47-51] adsorption than graphene due its buckled formation with a great deal of potential applications for silicene based nanoelectronic devices,[52] Li-ion storage batteries, hydrogen storage,[45] thin film solar cell absorbers,[46] hydrogen separation membranes,[47] and molecule sensors.[49-51]

Similar to graphene, silicene is a zero band gap semimetal.[15] It is expected that the effects of gas molecule adsorption on the electronic properties of a material without a band gap is much less than that of a semiconductor with an intrinsic band gap. A number of methods have been proposed to induce a band gap in a silicene sheet including doping,[53] substrate effects,[54] chemically functionalization,[55,56] electric field,[33] nanomesh and nanoholes.[57-59] One of the possible methods to introduce a band gap in silicene is achieved by cutting the sheet into silicene nanoribbons (SiNRs).[16,34,60] SiNRs have been grown on Ag (100), Ag (110), Au (110) substrates.[61-64] Afuray *et al.* have synthesized an array of highly uniform parallel SiNRs with width of ~ 1.7 nm on Ag (110) substrate.[61] The intrinsic band gap and small dimension along with free reactive edges make nanoribbons more attractive than sheets for gas

[a.] Quantum Electronic Structures Technology Lab, Department of Electrical and Computer Engineering, Florida International University, Miami, Florida 33174, United States
[b.] Department of Mechanical and Materials Engineering, Florida International University, Miami, Florida 33174, United States

nanosensor applications. Several experimental and theoretical studies of graphene nanoribbons (GNRs) sensing properties have been already reported.[65-68] The CO, NO, $NO_2$, and $O_2$ molecules are chemisorbed on armchair GNRs (AGNRs), while the adsorption of $NH_3$ and $CO_2$ on AGNRs is between weak chemisorption and strong physisorption.[67] It has been found that CO, $CO_2$, NO, $NO_2$, and $O_2$ molecules draw electrons from AGNRs, while $NH_3$ donates its electrons into AGNRs. Little attention has been focused on molecule adsorption on SiNRs. A recent theoretical work by Osborn and Farajian has proven that ASiNRs can be used for detection of CO molecule due to its weak chemisorption on ASiNRs.[69] They also found that $H_2O$ and $O_2$ are strongly chemisorbed on ASiNRs, while $CO_2$ and $N_2$ barely affect ASiNR's conductance.

It remains an open question as to what is the adsorption behavior of nitrogen-, sulfur-, and other carbon-based gas molecules, including NO, $NO_2$, $NH_3$, $SO_2$, $H_2S$, and $CH_4$ which are all of great practical interest for environmental, medical, and industrial applications, on SiNRs. In this paper, we employed first-principles methods based on density functional theory (DFT) to investigate the adsorption behavior of CO, $CO_2$, $CH_4$, $N_2$, NO, $NO_2$, $NH_3$, $SO_2$, and $H_2S$ molecules on SiNRs. Our results promulgate the promising future of SiNR in the development of ultrahigh sensitive sensor platforms.

## II Computational Methods

All the calculations are carried out based on first-principles DFT combined with nonequilibrium Green's function (NEGF) implemented in Atomistix ToolKit (ATK) package.[70-72] The Generalized Gradient Approximation of Perdew-Burke-Ernzerhof (GGA-PBE) with a double-ζ polarized basis set is adopted to solve Kohn-Sham equations and to expand electronic density. The density mesh cut off is set to be 150 Rydberg. The Grimme vdW correction (DFT-D2)[73] is also employed to describe long-range vdW interactions.[74] In order to take into account the vdW interactions, an additional term ($E_{vdW}$) is added to the DFT total energy ($E_{DFT}$):

$$E_{DFT-D2} = E_{DFT} + E_{vdW}$$

The $E_{vdW}$ is calculated using an attractive semi-empirical pair potential ($V^{PP}$):

$$E_{vdW} = 0.75 \sum_{m<m'} V^{PP}(Z_m, Z_{m'}, r_{m,m'})$$

The $V^{PP}$ between two atoms, for example atoms 1 and 2, which are located at a distance $r$ apart can be defined as:

$$V^{PP}(Z_1, Z_2, r_{12}) = \frac{\sqrt{C_6(Z_1)C_6(Z_2)}}{r_{12}^6} f(\frac{r_{12}}{R_0(Z_1)+R_0(Z_2)})$$

here $C_6$ and $R_0$ are the element-specific parameters and $f(x)$ is a cut-off function which is defined as:

$$f(x) = \frac{1}{1+e^{-dx}}$$

Typically, $d$ (damping parameter) is set to be 20. Besides vdW interactions, since our systems have two subsystems: the SiNR (A) and the gas molecule (B), so-called basis set superposition errors (BSSE) are expected due to the incompleteness of the Linear Combination of Atomic Orbitals (LCAO) basis set. In an isolated A system, only the basis orbitals in the A system are responsible to describe it. While, A and B are coupled, the basis orbitals in the system B will also be used to describe system A, resulting in a larger available basis set for system A. Consequently, there will be an artificial interaction which decreases the total energy. To eradicate BSSE, a counterpoise (cp) correction is added to the total energy:[75]

$$E_{BSSE} = E_{DFT} + E^{cp}$$

here $E_{DFT}$ is the total energy of the system AB and $E^{cp}$ is the counterpoise corrected energy which is:

$$E^{cp} = (E_A - E_{AB'}) + (E_B - E_{A'B})$$

where $E_{AB'}$ ($E_{A'B}$) is the energy of system A (B) using the AB basis orbitals, which is obtained by considering so-called ghost orbitals at the atomic positions in the system B (A). Ghost atoms have no charge and no mass; however, they have basis orbitals. $E_A$ ($E_B$) is the total energy of an isolated system A (B) using the A (B) basis orbitals. Finally, the total energy of the system considering the long-range vdW interactions and artificial attractions between two subsystems is:

$$E_{Total} = E_{DFT} + E_{vdW} + E^{cp}$$

To avoid the mirroring interactions, a vacuum space of 25 Å is considered in $x$ and $y$ directions in which the structures are not periodic. The electronic temperature is kept constant at 300 °K. All the structures are completely relaxed, prior to the calculations, up until the force and stress are less than 0.01 eV/Å and 0.005 eV/Å$^3$, respectively. 1×1×21 k-points in the Brillouin zone are sampled for geometry optimization and 1×1×121 k-points for total energy, band structure, charge transfer, and electron transport calculations.

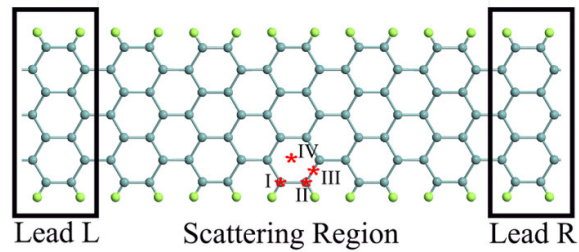

Fig. 1 Schematic structural model of SiNR-based gas sensor with two electrodes (black boxes). The cyan and green balls represent Si and H atoms, respectively. Possible adsorption sites of gas molecules on the SiNR are: I (hill), II (valley), III (bridge), and IV (hollow).

To investigate the charge transfer and transport properties, the gas sensing system is divided onto three regions: two electrode regions (left and right) and a scattering region (the central region), as illustrated in Fig. 1. To match the effective potential of central region with bulk electrodes, the perturbation of the scattering region should be screened out. To this end, a sufficient fraction of the electrode regions should be repeated in the scattering region. To calculate the non-equilibrium electron distribution in the central region, the NEGF method is employed. The charge density of the system based on the occupied eigenstates can be defined as

$$n(r) = \sum_a |\psi_a(r)|^2 f(\frac{\varepsilon_a - \varepsilon_f}{kT})$$

here $f(x)=1/(1+e^x)$ is the Fermi function, $\psi$ is the wave function, $\varepsilon_f$ is the Fermi energy, $T$ is the electron temperature, and $k$ is the Boltzmann constant. Conveniently, $n(r)$ can be presented in term of density matrix ($D_{ij}$)

$$n(r) = \sum_a D_{ij} f_i(r) f_j(r)$$

where $D_{ij}$ is defined by basis set expansion coefficients

$$D_{ij} = \sum_a c_{ai}^* c_{aj} f(\frac{\varepsilon_a - \varepsilon_f}{kT})$$

The density matrix is divided into left and right contributions

$$D = D^L + D^R$$

where $D^{L(R)}$ is calculated using NEGF theory by

$$D^{L(R)} = \int \rho^{L(R)}(\varepsilon) f(\frac{\varepsilon_a - \mu_{(R)}}{k_B T_{L(R)}}) d\varepsilon$$

Where $\rho^{L(R)}(\varepsilon)$, the spectral density matrix, is

$$\rho^{L(R)}(\varepsilon) \circ \frac{1}{2\pi} G(\varepsilon) \Gamma^{L(R)}(\varepsilon) G^\dagger(\varepsilon)$$

Here $\Gamma^{L(R)}$ is the broadening function of the left (right) electrode which defined as

$$\Gamma^{L(R)} = \frac{1}{i}(\Sigma^{L(R)} - (\Sigma^{L(R)})^\dagger)$$

where $\Sigma^{L(R)}$, the left (right) electrode self-energy, is

$$\Sigma^{L(R)} = V_{S/L(R)} g_{L(R)} V_{L(R)/S}$$

where $g_{L(R)}$ is the surface Green's function for the semi-infinite electrodes and $V_{L(R)/S} = V^\dagger_{S/(L/R)}$ are the coupling matrix elements between electrodes and the scattering region. Furthermore, the key quantity to calculate is $G$, the retarded Green's function matrix,

$$G(\varepsilon) = \frac{1}{(\varepsilon + id_+)S - H}$$

where $d_+$ is an infinitesimal positive number. $S$ and $H$ are the overlap and Hamiltonian matrices of the entire system, respectively. The Green's function is only required for the central region and can be calculated from the Hamiltonian of the central region by adding the electrode self-energies

$$G(\varepsilon) = [(\varepsilon + id_+)S - H - \Sigma^L(\varepsilon) - \Sigma^R(\varepsilon)]^{-1}$$

The transmission amplitude $t_k$, defines the fraction of a scattering state $k$ propagating through a device. The transmission coefficient at energy $\varepsilon$ is obtained by summing up the transmission from all the states at this energy,

$$T(\varepsilon) = \sum_k t_k^\dagger t_k \delta(\varepsilon - \varepsilon_k)$$

The transmission coefficient may also be obtained from the retarded Green's function using

$$T(\varepsilon) = G(\varepsilon) \Gamma^L(\varepsilon) G^\dagger(\varepsilon) \Gamma^R(\varepsilon)$$

At $T_R = T_L = 0$ (electron temperature) the conductance is determined by the transmission coefficient at the Fermi Level,

$$C(\varepsilon) = G_0 T(\varepsilon)$$

where $G_0 = 2e^2/h$ is the quantum conductance, in which $e$ is the electron charge and $h$ is Planck's constant.

## III Results and Discussions

In this study, all the nanosensors are made of long and non-periodic 7-ASiNRs. Their edges are passivated by H while their surfaces are kept pristine because the surfaces are less reactive than the edges.[76] It has been reported that 7-ASiNR is a paramagnetic semiconductor with a band gap of ~0.56 eV.[56,58] The semiconducting ASiNR is chosen because it is expected that the adsorption of gas molecules has a much smaller effect on the electronic properties of zigzag SiNRs (ZSiNRs) which are metallic. To evaluate chemical sensing of SiNR-based nanosensors, single gas molecules ($N_2$, NO, $NO_2$, $NH_3$, CO, $CO_2$, $CH_4$, $SO_2$, and $H_2S$) are initially placed about 3Å from the ASiNRs surface. The adsorption behaviors of molecules on SiNR are investigated after full relaxation. The structural stability of the molecule's adsorption on SiNR is addressed using adsorption Energy $(E_{ad})$ which is

$$E_{ad} = E_{SiNR+Molecule} - E_{SiNR} - E_{Molecule}$$

where $E_{SiNR+Molecule}$, $E_{SiNR}$, $E_{Molecule}$ denote the total energies of the SiNR-Molecule system, pristine SiNR, and the isolated gas molecule, respectively. Based on the definition, the more negative $E_{ad}$ is, the stronger adsorption of gas molecules on SiNR would be. By considering different adsorption configurations of gas molecules on SiNR and calculating each configuration's adsorption energy, it is discovered that different gas molecules prefer to be adsorbed with different geometries. To commence the relaxation, the molecules can be placed at four different positions including valley, hill, bridge sites, and hollow, as shown in Fig. 1. At these particular positions, different molecular orientations are considered. The structural stability of diatomic molecules (N$_2$, NO, and CO) are examined when their molecular axis is aligned perpendicular and parallel with respect to the SiNR's surface. Furthermore, for NO and CO molecules, the O atom can point up and down. For the triatomic molecule CO$_2$ with 180° bond angles, two orientations, parallel and perpendicular to the surface, are tested. For other triatomic (NO$_2$, SO$_2$, and H$_2$S), tetratomic (NH$_3$), and pentatonic (CH$_4$) molecules, two orientations are considered. In the first orientation, the N atom (in NO$_2$ and NH$_3$), S atom (in SO$_2$ and H$_2$S), and C atom (in CH$_4$) point to the SiNR's surface, while, in the second orientation, the H atom (in NH$_3$, H$_2$S, and CH$_4$), and O atom (in SO$_2$ and NO$_2$) point away from the SiNR's surface. The adsorption energies of gas molecules on the edge site and the interior site are also compared. We found that the gas molecules prefer to be adsorbed on the edge sites. Fig. 2 presents the most stable adsorption configurations of gas molecules on pristine ASiNRs.

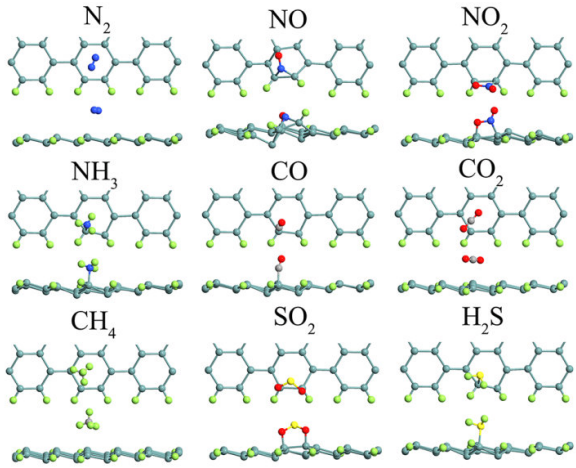

Fig. 2 The most stable adsorption configurations (top and side view) for N$_2$, NO, NO$_2$, NH$_3$, CO, CO$_2$, CH$_4$, SO$_2$, and H$_2$S on pristine ASiNR. The cyan, green, blue, red, grey, and yellow balls represent Si, H, N, O, C, and S atoms, respectively.

The calculated adsorption energies and binding distances between the gases and ASiNR are specified in Table 1. Our results show that CO is chemically adsorbed on SiNR with an adsorption energy of – 0.92 eV. In the most stable structure, the CO molecule prefers to be adsorbed in a vertical orientation when the C atom is connected to a Si-edge atom with a covalent bond at a hill position with a binding distance of 1.89 Å. The bond length of the adsorbed CO is 1.16 Å which is a little longer than that of an isolated CO molecule (1.14 Å). For CO$_2$ and CH$_4$, the results indicate that they are weakly physisorbed on SiNR with a small adsorption energy of – 0.31 and – 0.18 eV, respectively. Both molecules prefer to be placed at hollow site after relaxation. The N$_2$ molecule is energetically stable when it is horizontally situated at the center of a hexagonal silicon cell near the edge. The low adsorption energy of – 0.31 eV shows that N$_2$ is physically adsorbed on SiNRs. However, SiNR is highly reactive to other N-based gas molecules such as NH$_3$, NO$_2$, and NO, whose adsorption energies are – 1.12, – 2.53, – 2.68 eV, respectively.

NH$_3$ is chemically adsorbed on top of Si-edge atom at a hill position with the formation of a N-Si covalent bond (2 Å). The adsorption energy of NO$_2$ and NO is even less than – 2 eV showing a very strong chemisorption. For the NO$_2$ molecule, the N-O bond interacts with the Si-Si bond of SiNR at the edge, where the N-Si and O-Si covalent bonds' length are 1.91 and 1.77 Å, respectively. The chemisorption of the NO molecule on SiNR is even stronger than that of NO$_2$ molecule. The N atom is bonded to Si-Si atoms on the edge of the SiNR to form a N-Si-Si triangle with N-Si and Si-Si bond lengths of 1.80 and 2.32 Å, respectively. The O atom is also connected to another Si atom on top of the Si-hexagon with a bond length of 1.73 Å. The adsorption energy of H$_2$S molecule on top of SiNR is – 0.64 eV, showing that the adsorption of the molecule is physisorption. Strong chemisorption is also observed for the SO$_2$ molecule, whose adsorption energy is – 2.63 eV. In this case, the O atoms of SO$_2$ molecule are bonded to Si-Si bond at the SiNR edge, where the covalent bond lengths are 1.70 Å.

To sum up, the interaction of N$_2$, CO$_2$, CH$_4$, and H$_2$S gas molecules with a SiNR is mostly vdW type adsorbing *via* physisorption. A SiNR cannot be an appropriate sensor for detection of NO, NO$_2$, and SO$_2$ gas molecules because of their strong chemisorption on silicene. However, SiNR can be considered as a disposable molecule sensor for detection of NO, NO$_2$, and SO$_2$. Furthermore, the strong covalent bonding of these molecules to SiNR makes it possible to tune the electronic properties of SiNR for nanoelectronics applications, as will be discussed later. Finally, the moderate adsorption energies of NH$_3$ and CO gas molecules make SiNR a promising material for sensitive gas sensors or gas filters because they can be easily desorbed from SiNR by heating. The calculated adsorption energies of all studied gas molecules on SiNR are distinctly larger than those of silicene sheet and GNR,[49-51,67] showing higher sensitivity of SiNR toward the gas molecules. It should also be noted that environmental gas molecules (N$_2$, CO$_2$, O$_2$, and H$_2$O) can change the NH$_3$ and CO sensing capability SiNR's. Unlike physisorption of N$_2$ and CO$_2$ gas molecules on SiNR, the water and oxygen molecules are strongly chemisorbed on SiNR.[69] Therefore, it is critical that water and oxygen molecules should be removed from the environment to preserve the proper sensing capability of SiNRs.

To investigate the effects of gas molecules on the conductance of SiNR, the quantum conductances of the pristine SiNR before and after adsorption are calculated, as shown in Fig. 3. Upon adsorption of N-based gas molecules, the band gap of pristine SiNR (0.56 eV) is decreased slightly for N$_2$ (0.52 eV) and NH$_3$ (0.48 eV) and is somewhat increased for NO$_2$ (0.68 eV) and NO (0.68 eV). The conductances of SiNR before and after adsorption of N$_2$ molecule are similar due to the physisorption of N$_2$ gas molecule on SiNR, showing insensitivity of SiNR sensor to N$_2$ gas molecule. However, the overall conductances of SiNRs are detectably dropped after NH$_3$, NO$_2$, and NO, confirming their chemisorption on SiNRs. The conductance reduction for a NO molecule is more vivid than other N-based gas molecules which are consistent with the calculated adsorption energies. For C-based gas molecules, the band gap of pristine SiNR is preserved for $CO_2$ (0.56 eV) and slightly decreased for $CH_4$ (0.52 eV) and CO (0.48 eV). The overall conductance of pristine SiNR is visibly reduced after CO adsorption, while barely changed after $CO_2$ adsorption, and almost unchanged after $CH_4$ adsorption which confirms the calculated adsorption energies for C-based gas molecules where the CO adsorption energy is more negative than others. Upon adsorption of S-based gas molecules, the band gap of pristine SiNR is unaltered for $SO_2$ (0.56 eV) and is diminished for $H_2S$ (0.48 eV) gas molecule. A reduction in the conductance of the SiNR from $SO_2$ and $H_2S$ adsorption is also observed, confirming the sensing capability of SiNR for $SO_2$ and $H_2S$ gas molecules.

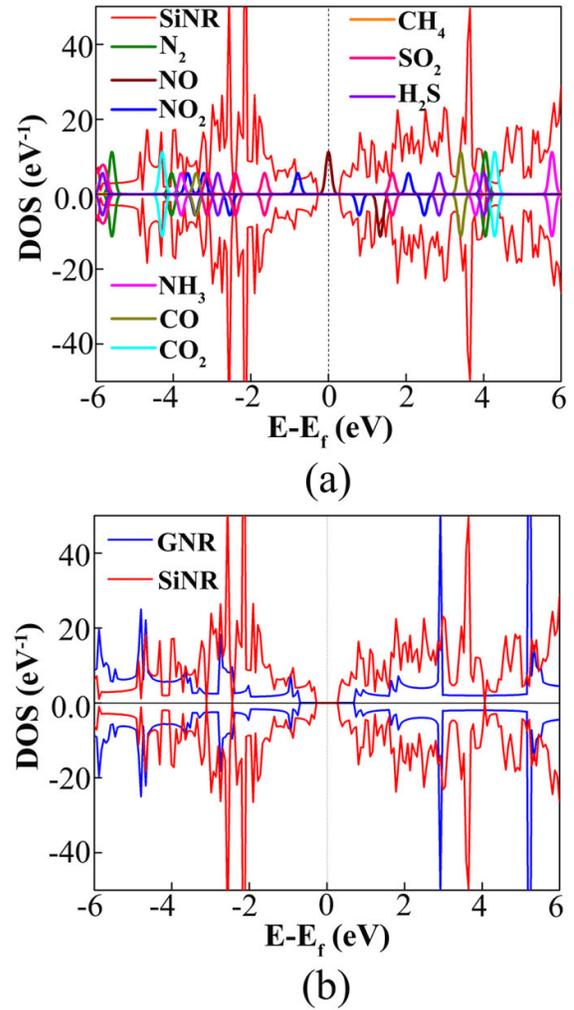

Fig. 4 (a) DOS of $N_2$, NO, $NO_2$, $NH_3$, b) CO, $CO_2$, $CH_4$, $SO_2$ and $H_2S$ gas molecules and SiNR. The positive and negative values represent spin-up and -down states, respectively. (b) DOS of SiNR and GNR. The Fermi level of SiNR and GNR are set to zero.

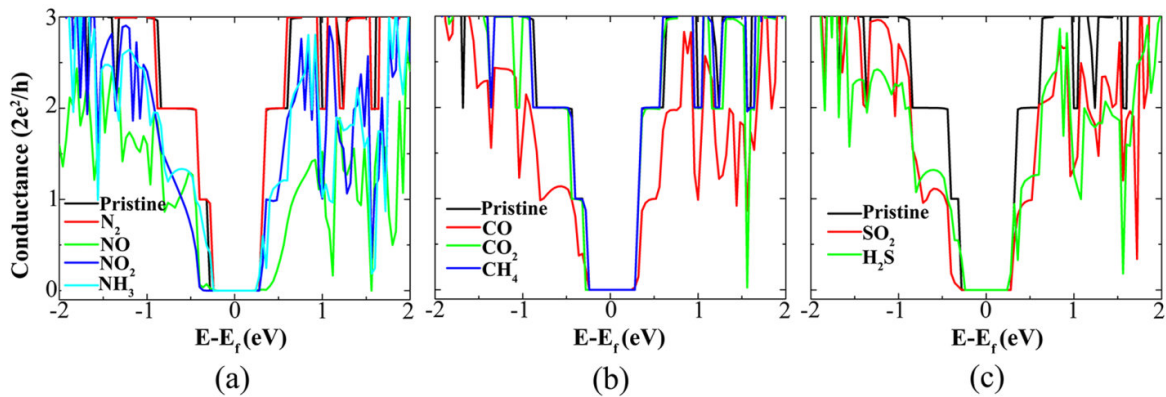

Fig. 3 Quantum conductance of pristine SiNR before and after (a) $N_2$, NO, $NO_2$, $NH_3$, (b) CO, $CO_2$, $CH_4$, (c) $SO_2$ and $H_2S$ gas molecules adsorption.

Table 1 The calculated adsorption energy ($E_{ad}$), binding distance which is the shortest atom to atom distance between molecule and device ($D$), and the charge transfer from ribbon to molecule ($r$)

| Device | Gas | $E_{ad}$ (ev) | D (Å) | r (e) | Device | Gas | $E_{ad}$ (ev) | D (Å) | r (e) |
|---|---|---|---|---|---|---|---|---|---|
| P-SiNR | $N_2$ | −0.30 | 3.47 | −0.110 | DB-SiNR | $N_2$ | −0.32 | 3.28 | −0.090 |
| | NO | −2.68 | 1.73 | −0.892 | | NO | −3.23 | 1.74 | −0.908 |
| | $NO_2$ | −2.53 | 1.77 | −0.851 | | $NO_2$ | −2.94 | 1.92 | −0.848 |
| | $NH_3$ | −1.12 | 2.00 | −0.535 | | $NH_3$ | −1.24 | 1.98 | −0.520 |
| | CO | −0.92 | 1.89 | −0.406 | | CO | −1.52 | 1.83 | −0.395 |
| | $CO_2$ | −0.31 | 3.34 | −0.101 | | $CO_2$ | −0.35 | 3.27 | −0.100 |
| | $CH_4$ | −0.18 | 3.40 | +0.025 | | $CH_4$ | −0.22 | 3.33 | +0.020 |
| | $SO_2$ | −2.64 | 1.74 | +0.626 | | $SO_2$ | −2.72 | 1.80 | −0.615 |
| | $H_2S$ | −0.64 | 2.45 | −0.342 | | $H_2S$ | −0.79 | 2.45 | −0.323 |
| Device | Gas | $E_{ad}$ (ev) | D (Å) | r (e) | Device | Gas | $E_{ad}$ (ev) | D (Å) | r (e) |
| N-SiNR | $N_2$ | −0.35 | 2.59 | −0.151 | B-SiNR | $N_2$ | −0.55 | 1.46 | −0.352 |
| | NO | −3.74 | 1.72 | +0.138 | | NO | −3.82 | 1.38 | −0.233 |
| | $NO_2$ | −3.60 | 1.75 | −0.041 | | $NO_2$ | −3.53 | 1.55 | −0.162 |
| | $NH_3$ | −1.35 | 1.98 | −0.559 | | $NH_3$ | −1.14 | 1.63 | −0.663 |
| | CO | −1.89 | 2.00 | −0.274 | | CO | −1.59 | 1.50 | −0.452 |
| | $CO_2$ | −0.35 | 2.70 | −0.140 | | $CO_2$ | −0.32 | 3.28 | −0.110 |
| | $CH_4$ | −0.27 | 3.57 | +0.016 | | $CH_4$ | −0.25 | 3.39 | +0.014 |
| | $SO_2$ | −3.48 | 1.75 | −0.210 | | $SO_2$ | −2.95 | 1.57 | −0.204 |
| | $H_2S$ | −1.86 | 2.40 | −0.305 | | $H_2S$ | −0.90 | 1.98 | −0.407 |

The electron charge transfer between gas molecules and SiNR is also investigated using Mulliken population analysis, as shown in Table 1. The positive value of charge indicates a charge transfer from SiNR to molecule. $N_2$, NO, $NO_2$, $NH_3$, CO, $CO_2$, and $H_2S$ are electron-donating gas molecules, while $CH_4$ and $SO_2$ have an electron-withdrawing capability. A large amount of charge transfer is observed for NO (−0.892 $e$), $NO_2$ (−0.851 $e$), $NH_3$ (−0.535 $e$) and CO (−0.406 $e$) due to their strong electron-donating characteristics. These large charge transfer molecules to SiNR are also correlated to the strong binding energies of the gas molecules chemisorption on SiNR.

The origin of the chemisorption of NO, $NO_2$, $NH_3$, CO, and $SO_2$ and physisorption of $N_2$, $CO_2$, $CH_4$, and $H_2S$ of gas molecules on SiNR can be revealed by studying their density of states (DOS). Fig. 4(a) presents the total DOS of pristine SiNR and the gas molecules. The frontier molecular orbitals, the highest occupied molecular orbital (HOMO) and the lowest unoccupied molecular orbital (LUMO), of NO, $NO_2$, $NH_3$, CO, and $SO_2$ are close to the Fermi level of SiNR. It shows that they have higher reactivity to SiNR as compared to $N_2$, $CO_2$, $CH_4$, and $H_2S$ which are physisorbed on the SiNR. These results are consistent with earlier research on gas molecule adsorption on silicene sheet.[48-51] As previously mentioned, silicene is more reactive than graphene because of its tendency to adopt $sp^2$ + $sp^3$ hybridization over $sp^2$ hybridization.[22] The difference between chemical reactivities of silicene and graphene can be understood from Fig. 4(b). It shows that the corresponding electronic states of 7-ASiNR are closer than that of 7-AGNR to the Fermi level, causing stronger reactivity of SiNR for molecule adsorption.

Since upon adsorption of individual gas molecules a conduction reduction is observed in the properties of SiNRs, significant alterations in the electronic properties of SiNRs are expected at higher concentrations of gas molecules. To evaluate this premise, a higher surface coverage with two gas molecules is considered and the impacts of gas molecules adsorption on the conduction of SiNR are investigated. Fig. 5 presents the most stable configurations of two adsorbed gas molecules on opposite edges of the SiNR. Based on the calculations, an individual gas molecule can be adsorbed anywhere at the SiNR's edges and its position will not change the quantum conductance. The second gas molecule can also be adsorbed on the same side or the opposite side of the first gas molecule. The results show that the conductance changes are almost the same for two cases, in agreement with the findings of Osborn et al.[69]

gas adsorption above the DB defect is tested. The most stable adsorption configuration of gas molecules on SiNR with an edge DB defect are illustrated in Fig. 7.

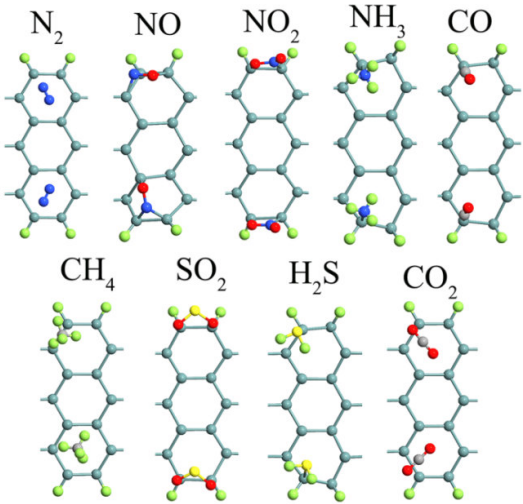

Fig. 5 The most stable adsorption configurations for two $N_2$, NO, $NO_2$, $NH_3$, CO, $CO_2$, $CH_4$, $SO_2$, and $H_2S$ on pristine ASiNR.

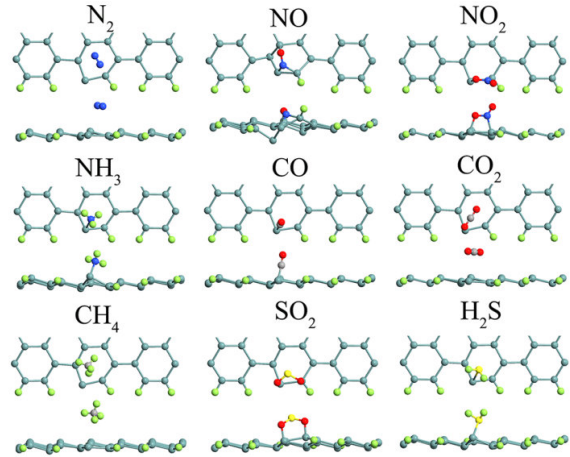

Fig. 7 The most stable adsorption configurations for $N_2$, NO, $NO_2$, $NH_3$, CO, $CO_2$, $CH_4$, $SO_2$, and $H_2S$ on ASiNR with an edge DB defect.

By calculating adsorption energies, it is found that adding an additional gas molecule to the system doubles the value of adsorption energies of gas molecules on the SiNR. It is clear that the NO, $NO_2$, $NH_3$, CO, $SO_2$, and $H_2S$ sensing capability of SiNR, which is exposed by two gas molecules, is increased due to the significant reduction in the conductance of pristine SiNR and the strong chemisorption of gas molecules to the SiNR (their adsorption energies are less than −1 eV), as shown in Fig. 6. By performing Mulliken population analysis of the SiNR with one adsorbed molecule and comparing to the SiNR with two adsorbed molecules, it is found that there are no appreciable differences between the amounts of charge transfer.

It is predicted that the edges of SiNRs are not well controlled similar to GNRs.[77,78] As a result, it seems difficult to achieve a fully edge hydrogenated SiNR without any dangling bond (DB) defects. It is found that edge DB defects are much more probable than surface DB defects in silicon nanowires.[79-82] Therefore, it is necessary to investigate the sensing capability of a SiNR with an edge DB defect. To this end, a hydrogen atom is removed from one silicone edge atom, and only the

Fig. 8 shows the conductance modulation of ASiNR with an edge DB defect upon adsorption of gas molecules. As one can see the edge DB defects do not limit the sensing capability of the SiNR. The absolute value of adsorption energies of all gas molecules on SiNR with an edge DB defect, see Table 1, are increased, leading to detectable modifications in SiNR's conductance. Based on their adsorption energies, it can be understood that NO (− 3.23 eV), $NO_2$ (− 2.94 eV), $NH_3$ (− 1.24 eV), CO (− 1.52 eV), and $SO_2$ (− 2.72 eV) are strongly chemisorbed on the SiNR with a DB defect, while $H_2S$ (− 0.79 eV) is strongly physisorbed on the SiNR. However, the $N_2$ (− 0.32 eV) and $CH_4$ (− 0.22 eV) molecules are still weakly physisorbed on the SiNR with small adsorption energy. The Mulliken population analysis of the SiNR with an edge DB defect and adsorbed molecule shows that a DB defect slightly decreases the charge transfer between SiNR and gas molecules, see Table 1.

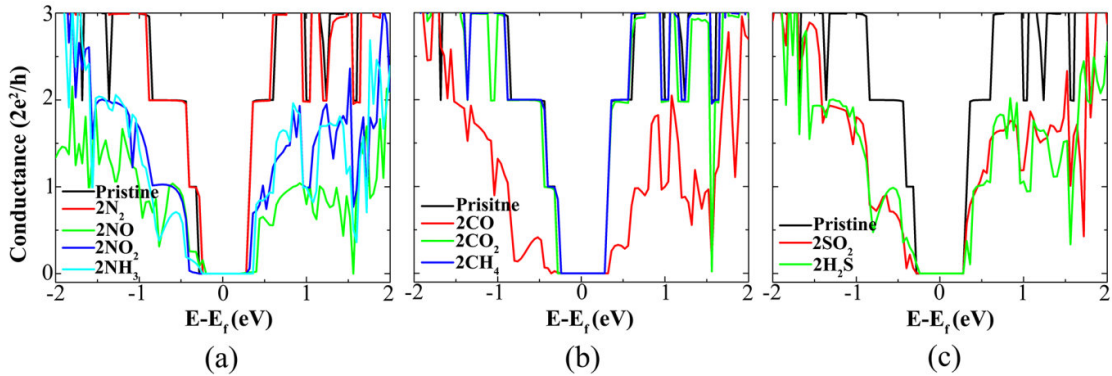

Fig. 6 Quantum conductance of pristine SiNR before and after a) two-$N_2$, -NO, -$NO_2$, -$NH_3$, b) -CO, -$CO_2$, -$CH_4$, c) -$SO_2$ and -$H_2S$ gas molecules adsorption.

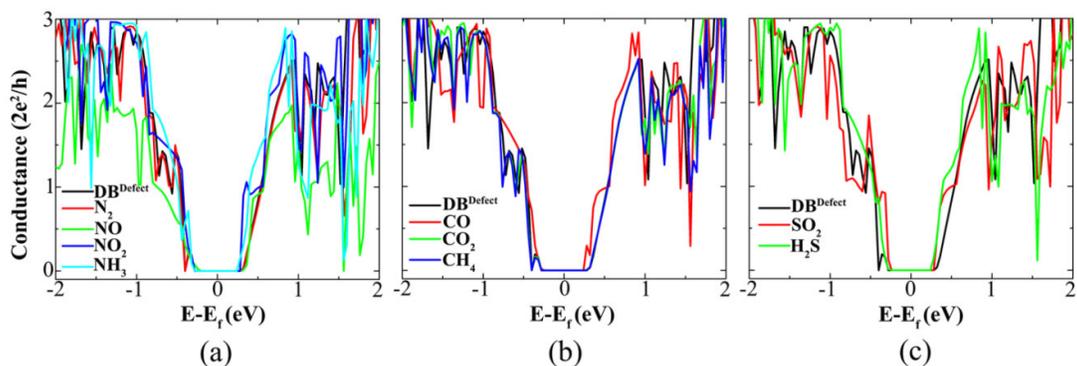

Fig. 8 Quantum conductance of SiNR with an edge DB defect before and after a) $N_2$, NO, $NO_2$, $NH_3$, b) CO, $CO_2$, $CH_4$, c) $SO_2$ and $H_2S$ gas molecules adsorption.

Here, we study the effects of doping on the gas molecules sensing capability of SiNR. To this end, doping effects of SiNRs with N and B impurities have been considered. It has been found that a single N or B dopant prefers to be substituted with Si edge atoms because the formation energy of N or B impurity at the edge is lower than other positions of the nanoribbon.[56] To focus on the effects of doping on the sensing capability of the SiNR, the gas molecules adsorption above the B and N impurities and their nearest neighbours are studied. It is discovered that the gas molecules energetically tend to be chemisorbed on B atom in B-doped SiNR, and chemisorbed on the Si atoms nearest to the N atom in N-doped SiNR. The adsorption energies of gas molecules on the doped SiNR are larger than that of pristine SiNR, proving that doping improves sensing capabilities of SiNR, see Table 1. These findings are in a good agreement with previous studies on gas detection based on doped silicene.[50]

The most stable configuration of adsorbed gas molecules on N-doped and B-doped ASiNR are depicted in Fig. 9. For CO gas molecule on N-doped SiNR, two covalent bonds are formed between C and two nearest Si atoms to the N atom, while, for B-doped SiNR, C atom and B atom are covalently bonded. The adsorption energies of CO molecule on N- and B-doped SiNR are 2.04 and 1.72, greater than that of pristine SiNR. The NO gas molecule is also strongly chemisorbed on N- and B-doped SiNR. Although three covalent bonds are formed between NO and nearest Si atoms to the N atom in N-doped SiNR, the adsorption of NO on B-doped SiNR totally destroys the configuration of atoms on the edge. The most stable configurations for $NO_2$, $NH_3$, and $SO_2$ gas molecules adsorption on N- and B-doped SiNR are quite similar to the pristine SiNR. The only difference is that they prefer to be adsorbed on top of a B atom and the nearest Si atom to the N atom. The $CO_2$ and $CH_4$ molecules on the N- and B-doped SiNR and $N_2$ and $H_2S$ on the N-doped SiNR are still physisorbed. However, $N_2$ and $H_2S$ gas molecules are chemically adsorbed on B-doped SiNR with adsorption energies of − 0.55 eV and − 0.90 eV which are 1.79 and 1.40 times greater than that of pristine SiNR. Similar to gas adsorption on pristine SiNR, the conductances of N-doped and B-doped SiNR are detectably changed upon adsorption of NO, $NO_2$, $NH_3$, CO, $SO_2$, and $H_2S$, see Fig. 10. Furthermore, doping a B atom into the SiNR can improve the $N_2$ gas sensing capability of SiNR, compared to that of pristine SiNR. Charge transfer analysis shows that $N_2$, $NO_2$, $NH_3$, CO, $CO_2$, $H_2S$, and $SO_2$ act as donors, while NO and $CH_4$ act as acceptors for N-doped SiNR. However, for B-doped SiNR, all gases have electron donating capability except $CH_4$, see Table 1.

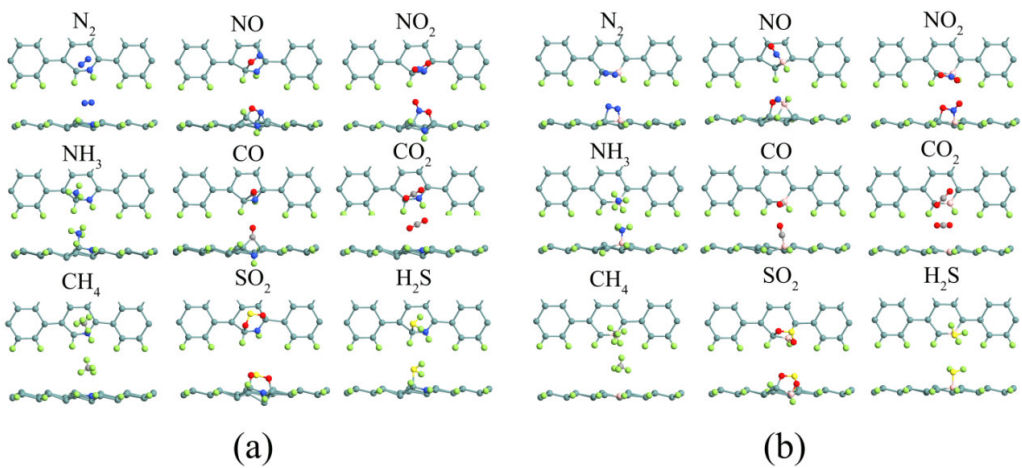

Fig. 9 The most stable adsorption configurations for $N_2$, NO, $NO_2$, $NH_3$, CO, $CO_2$, $CH_4$, $SO_2$, and $H_2S$ on (a) N-doped and (b) B-doped ASiNR. The cyan, green, blue, red, grey, yellow, and pink balls represent Si, H, N, O, C, S, and B atoms, respectively.

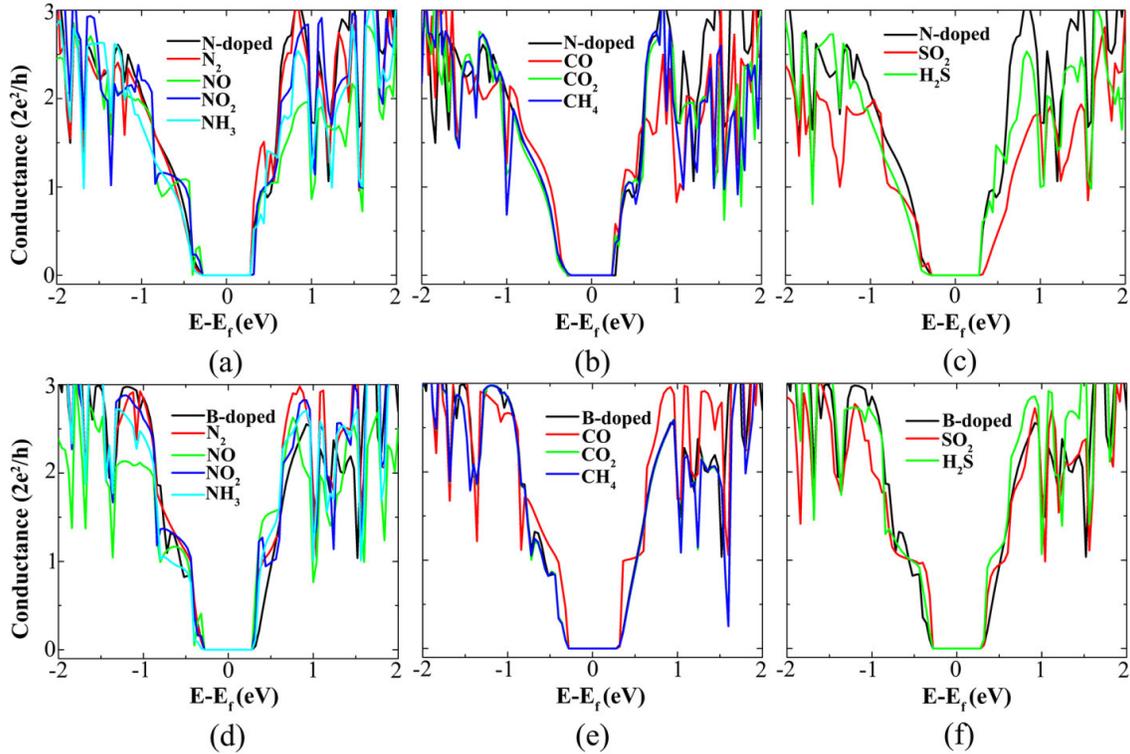

Fig. 10 Quantum conductance of N-doped SiNR before and after (a) $N_2$, NO, $NO_2$, $NH_3$, (b) CO, $CO_2$, $CH_4$, (c) $SO_2$ and $H_2S$ gas molecules adsorption. Quantum conductance of B-doped SiNR before and after (a) $N_2$, NO, $NO_2$, $NH_3$, (b) CO, $CO_2$, $CH_4$, (c) $SO_2$ and $H_2S$ gas molecules adsorption

Finally, a design of SiNR-based sensor to detect gas molecules is proposed, as shown in the inset of Fig. 11 (a). Two semi-infinite 4-ZSiNRs as the leads and a 98 Å long 7-ASiNR gas sensing zone are the main elements of the device. We have assumed that 5 gas molecules with an average distance of 18 Å are adsorbed on the edge sites of the detection zone. By applying a bias voltage across the leads, conductance can be measured before and after adsorption. To assess the sensing performance of the proposed device for the gas molecules the current *versus* bias voltage is calculated in Fig. 11 (a). For clarity and brevity, the $I$-$V_{bias}$ curves for 7-ASiNR before and after adsorption of $NO_2$ and NO molecules are considered because the currents induced by other molecules are much smaller than that of $NO_2$ and NO molecules. As can be seen, the current is always almost zero for pristine 7-ASiNR before adsorption of molecules since 7-ASiNR is semiconductor with a direct band gap of 0.56 eV, as shown in Fig. 11 (b). However, upon $NO_2$ adsorption, the current remarkably increases as a function of increasing voltage bias, the curve is almost linearindicating a metallic behavior. The current can be also increased by NO adsorption; however, the current value is much smaller than that of $NO_2$. These results can be confirmed by analyzing the band structures of 7-ASiNR after NO and $NO_2$ adsorption, as shown in Fig. 11 (b). As mentioned before, the gas molecules are able to tune the electronic properties of SiNRs. The adsorption of $NO_2$ and NO on 7-ASiNR can transform the system to n-type and p-type semiconductors, respectively, due to the deep defect states induced by gas molecules, as shown in Fig. 11 (b). These n- and p-type semiconducting SiNRs can be applied in nanoscale electronic devices such as a p-n junction diode and n- or p-type FETs. It is also expected that $NO_2$ gas molecule adsorption by SiNR can give rise to a conductance enhancement because of the metallic behavior of SiNR after $NO_2$ molecule adsorption. However, for NO molecule adsorption, the conductance may be enhanced by some amount since the NO adsorption makes the SiNR a p-type semiconductor.

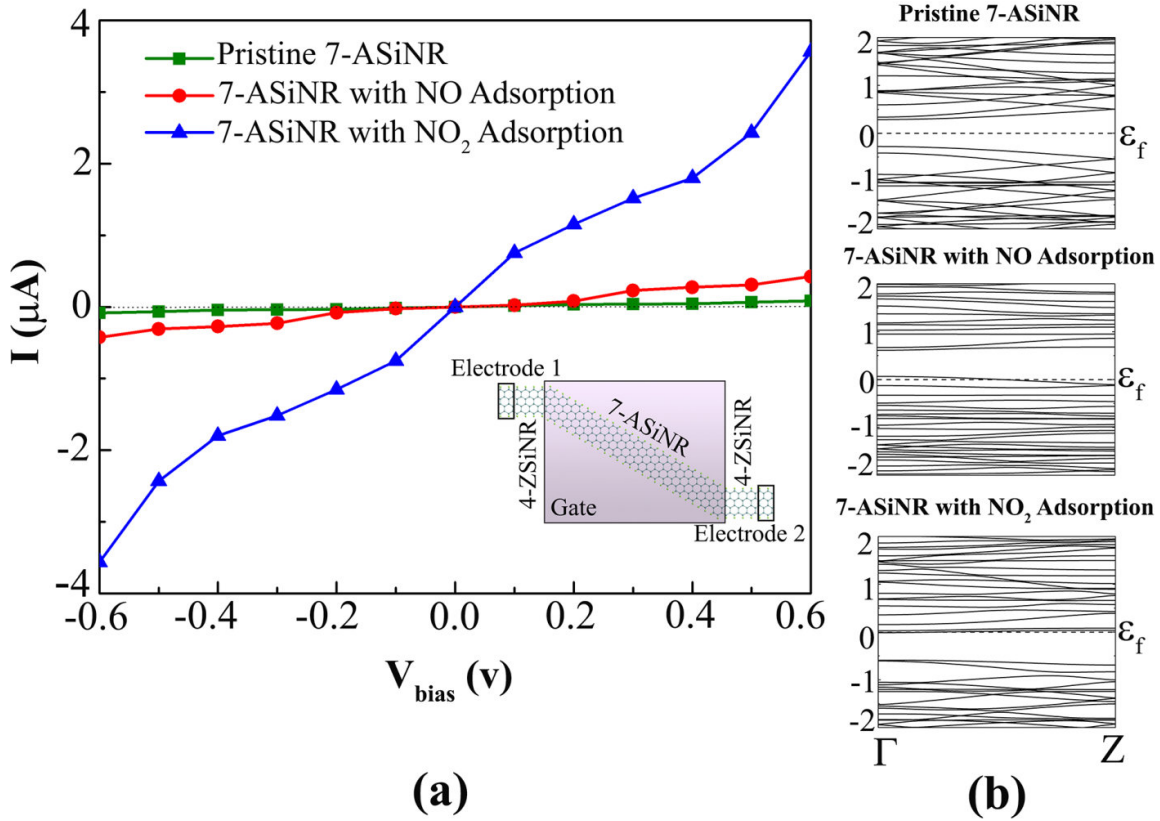

Fig. 11 (a) $I$-$V_{bias}$ curves for the SiNR sensor before and after NO and $NO_2$ adsorption. The inset shows the schematic of a SiNR-based sensor which consists of a semiconducting 7-ASiNR (detection region) and two metallic 4-ZSiNRs (electrodes). (b) Band structures of pristine 7-ASiNR before and after NO and $NO_2$ adsorption.

## IV Conclusions

In summary, we employed first-principle calculations to explore the adsorption geometry, adsorption energy, charge transfer, electronic band structure, and quantum conductance modulation of SiNRs with gas molecules ($N_2$, NO, $NO_2$, $NH_3$, CO, $CO_2$, $CH_4$, $SO_2$, and $H_2S$) adsorption. Our results reveal that SiNRs are capable of detecting CO and $NH_3$ with high sensitivity because they are chemically adsorbed on the SiNR and transfer a few electrons to the SiNR. Quantum conductance modulation of SiNRs is clearly detectable upon chemisorption of gas molecules. Furthermore, SiNRs are not appropriate to serve as a sensor for NO, $NO_2$, and $SO_2$ due to the fact that they are strongly chemisorbed with covalent bonds to SiNRs and transfer a large amount of electrons to the SiNR. However, the strong bonding of NO and $NO_2$ molecules is an effective way to tune the electronic properties of SiNRs to fabricate p-type and n-type transistors, respectively. The other gas molecules are physisorbed. We also found that increasing the density of gas molecules will result in more significant changes in quantum conductance. In addition, dangling bond defects which are unavoidable through fabrication process will not hinder the sensing capability of SiNRs. The sensing capability of a SiNR-based sensor can also be increased by either N or B doping. It is found that doping a SiNR with a B atom can enhance the detection capability of $N_2$ gas molecules. On the basis of our results, SiNRs can be considered as a promising material to detect individual gas molecules.

## Acknowledgements

This work was supported in part by the Florida Education Fund's McKnight Junior Faculty Fellowship.

## References

1. K. S. Novoselov, A. K. Geim, S. Morozov, D. Jiang, Y. Zhang, S. Dubonos, I. Grigorieva, and A. Firsov, *Science*, 2004, **306**, 666-669.
2. K. Novoselov, A. K. Geim, S. Morozov, D. Jiang, M. Katsnelson, I. Grigorieva, S. Dubonos, and A. Firsov, *Nature*, 2005, **438**, 197-200.
3. S. Basu and P. Bhattacharyya, *Sens. Actuators B,* 2012, **173**, 1-21.
4. F. Schedin, A. K. Geim, S. V. Morozov, E. W. Hill, P. Blake, M. I. Katsnelson, and K. S. Novoselov, *Nat. Mater.*, 2007, **6**, 652–655.

5. Q. He, S. Wu, Z. Yin, H. Zhang, *Chem. Sci.*, 2012, **3**, 1764-1772.
6. J. D. Fowler, M. J. Allen, V. C. Tung, Y. Yang, R. B. Kaner, and B. H. Weiller, *ACS Nano,* 2009, **3**, 301-306.
7. S. Some, Y. Xu, Y. Kim, Y. Yoon, H. Qin, A. Kulkarni, T. Kim, H. Lee, *Sci. Rep.,* 2013, **3**, 1868.
8. O. Leenaerts, B. Partoens, and F. M. Peeters, *Phys. Rev. B*, 2008, **77**, 125416.
9. Y. Zhang, Y. Chen, K. Zhou, C. Liu, J. Zeng, H. Zhang, and Y. Peng, *Nanotechnology*, 2009, **20**, 185504.
10. J. Dai, J. Yuan, and P. Giannozzi, *Appl. Phys. Lett*., 2009, **95**, 232105.
11. J. T. Robinson, F. K. Perkins, E. S. Snow, Z. Q. Wei, and P. E. Sheehan, *Nano Lett*., 2008, **8**, 3137.
12. G. Lu, L. E. Ocola, and J. Chen, *Nanotechnology*, 2009, **20**, 445502.
13. J. Dai and J. Yuan, *Phys. Rev. B*, 2010, **81**, 165414.
14. K. Takeda and K. Shiraishi, *Phys. Rev. B*, 1994, **50**, 14916.
15. G. G. Guzmán-Verri and L. L. Y. Voon, *Phys. Rev. B*, **2007**, 76, 075131.
16. Cahangirov, M. Topsakal, E. Aktürk, H. Şahin and S. Ciraci, *Phys. Rev. Lett.*, 2009, **102**, 236804.
17. X. Q. Wang, H. D. Li, and J. T. Wang, *Phys. Chem. Chem. Phys.*, 2012, **14**, 3031–3036.
18. F. Zheng and C. Zhang, *Nanoscale Res. Lett.*, 2012, **7**, 422.
19. C.C. Liu, W.X. Feng, and Y.G. Yao, *Phys. Rev. Lett.*, 2011, **107**, 076802.
20. C. Xu, G. Luo, Q. Liu, J. Zheng, Z. Zhang, S. Nagase, Z. Gao, and J. Lu, *Nanoscale*, 2012, **4**, 3111-3317.
21. L. Chen, B. Feng, and K. Wu, *Appl. Phys. Lett*., 2013, **102**, 081602.
22. P. Vogt, P. De Padova, C. Quaresima, J. Avila, E. Frantzeskakis, M. C. Asensio, A. Resta, B. Ealet, and G. Le Lay, *Phys. Rev. Lett.*, 2012, **108**, 155501.
23. B. Feng, Z. Ding, S. Meng, Y. Yao, X. He, P. Cheng, L. Chen, and K. Wu, *Nano Lett.,* 2012, **12**, 3507–3511.
24. J. Mannix, B. Kiraly, B. L. Fisher, M. C. Hersam, and N. P. Guisinger, *ACS Nano*, 2014, **8**, 7538-7547.
25. L. Meng, Y. Wang, L. Zhang, S. Du, R. Wu, L. Li, Y. Zhang, G. Li, H. Zhou, W. A. Hofer *et al*., *Nano Lett.*, 2013, **13**, 685-690.
26. A. Fleurence, R. Friedlein, T. Ozaki, H. Kawai, Y. Wang, and Y. Yamada-Takamura, *Phys. Rev. Lett.* 2012, **108**, 245501.
27. T. Aizawa, S. Suehara, and S. Otani, *J. Phys. Chem. C*, 2014, **118**, 23049-23057.
28. T. P. Kaloni, M. Tahir, and U. Schwingenschlögl, *Sci. Rep*. 2013, **3**, 3192.
29. H. Liu, J. Gao, and J. Zhao, *J. Phys. Chem. C*, 2013, 117, 10353.
30. Y. Ding and Y. Wang, *Appl. Phys. Lett*., 2013, **103**, 043114.
31. Y. Cai, C.-P. Chuu, C. M. Wei, and M. Y. Chou, *Phys. Rev. B*, 2013, **88**, 245408.
32. M. Houssa, B. van den Broek, E. Scalise, G. Pourtois, V. V. Afanas'eva and A. Stesmansa, 2013, *Phys. Chem. Chem. Phys.* 15, 3702.
33. W.-F. Tsai, C.-Y. Huang, T.-R. Chang, H. Lin, H.-T. Jeng and A. Bansil, *Nat. Commun.*, 2013, **4**, 1500.
34. Z. Ni, Q. Liu, K. Tang, J. Zheng, J. Zhou, R. Qin, Z. Gao, D. Yu, and J. Lu, *Nano Lett.* 2012, **12**, 113-118.
35. L. Tao, E. Cinquanta, D. Chiappe, C. Grazianetti, M. Fanciulli, M. Dubey, A. Molle, and D. Akinwande, *Nat. Nanotechnol.*, 2015, **10**, 227-231.
36. H. Sadeghi, S. Bailey, C. J. Lambert, *Appl. Phys. Lett.,* 2014, **104**, 103104.
37. R. G. Amorim and R. H. Scheicher, *Nanotechnology*, 2015, **26**, 154002.
38. N. D. Drummond, V. Zólyomi, and V. I. Fal'ko, *Phys. Rev. B*, 2012, **85**, 075423.
39. T. P. Kaloni, G. Schreckenbach, and M. S. Freund*, J. Phys. Chem. C,* 2014, **118**, 23361-23367.
40. P. Lazar, F. Karlický, P. Jurečka, M. Kocman, E. Otyepková, K. Šafářová, and M.; Otyepka, *J. Am. Chem. Soc*. 2013, **135**, 6372–6377.
41. X. Lin and J. Ni, *Phys. Rev. B*, 2012, 86, 075440.
42. J. Sivek, H. Sahin, B. Partoens, F. M. Peeters, *Phys. Rev. B,* 2013, **87**, 085444.
43. H. Sahin and F. M. Peeters, *Phys. Rev. B,* 2013, **87**, 085423.
44. G. A. Tritsaris, E. Kaxiras, S. Meng, and E. Wang, *Nano Lett.*, 2013, **13**, 2258-2263.
45. J. Wang, J. Li, S.-S. Li and Y. Liu, *J. Appl. Phys.*, 2013, **114**, 124309.
46. B. Huang, H. J. Xiang and S.-H. Wei, *Phys. Rev. Lett.,* 2013, **111**, 145502.
47. W. Hu, X. Wu, Z. Li and J. Yang, *Phys. Chem. Chem. Phys*., 2013, **15**, 5753.
48. J.-W. Feng, Y.-J. Liu, H.-X. Wang, J.-X. Zhao, Q.-H. Cai and X.-Z. Wang, *Comput. Mater. Sci.,* 2014, **87**, 218.
49. W. Hu, N. Xia, X. Wu, Z. Li and J. Yang, *Phys. Chem. Chem. Phys.,* 2014, **16**, 6957.
50. J. Prasongkit, R. G. Amorim, S. Chakraborty, R. Ahuja, R. H. Scheicher, and V. Amornkitbamrung, *J. Phys. Chem. C*, 2015, **119**, 16934-16940.
51. R. Chandiramoulia, A. Srivastavab, and V. Nagarajan, *Appl. Surf. Sci.*, 2015, **351**, 662-672.
52. A. Kara, H. Enriquez, A. P. Seitsonen, L. C. Lew Yan Voon, S. Vizzini, B. Aufray and H. Oughaddou, *Surf. Sci. Rep*., 2012, **67**, 1-18.
53. A. Lopez-Bezanilla, *J. Phys. Chem. C*, 2014, **118**, 18788-18792.
54. N. Gao, J. Li, and Q. Jiang, *Chem. Phys. Lett.*, 2014, **592**, 222-226.
55. N. Gao, W. T. Zheng and Q. Jiang, *Phys. Chem. Chem. Phys.*, 2012, **14**, 257-261.
56. S. M. Aghaei, M. M. Monshi, I. Torres, and I. Calizo, *RSC Adv*., 2016, **6**, 17046.
57. F. Pan, Y. Wang, K. Jiang, Z. Ni, J. Ma, J. Zheng, R. Quhe, J. Shi, J. Yang, C. Chen, and J. Lu, *Sci. Rep*. 2015, **5**, 9075.
58. S. M. Aghaei and I. Calizo, *J. Appl. Phys.,* 2015, **118**, 104304.
59. S. M. Aghaei and I. Calizo, in Proceeding of IEEE SoutheastCon (SECon-2015), Fort Lauderdale, April 9-12, 2015, pp. 1-6.
60. Y. Ding and J. Ni, *Appl. Phys. Lett*., 2009, **95**, 083115.
61. B. Aufray, A. Kara, S. Vizzini, H. Oughaddou, C. Leandri, B. Ealet, and G. Le Lay, *Appl. Phys. Lett*., 2010, **96**, 183102.
62. P. De Padova, C. Quaresima, C. Ottaviani, P. M. Sheverdyaeva, P. Moras, C. Carbone, D. Topwal, B. Olivieri, A. Kara, H. Oughaddou, B. Aufray, G. Le Lay*, Appl. Phys. Lett*., 2010, **96**, 261905.
63. P. De Padova, O. Kubo, B. Olivieri, C. Quaresima, T. Nakayama, M. Aono, and G. Le Lay, *Nano Lett*., 2012, **12**, 5500-5503.
64. M. R. Tchalala, H. Enriquez, A. J. Mayne, A. Kara, S. Roth, M. G. Silly, A. Bendounan, F. Sirotti, T. Greber, B. Aufray, G. Dujardin, M. A. Ali, and H. Oughaddou*, Appl. Phys. Lett*., 2013, **102**, 083107.
65. A. N. Abbas, G. Liu, B. Liu, L. Zhang, H. Liu, D. Ohlberg, W. Wu and C. Zhou, *ACS Nano*, 2014, **8**, 1538-1546.
66. A. N. Abbas, G. Liu, A. Narita, M. Orosco, X. Feng, K. Müllen and C. Zhou, *J. Am. Chem. Soc.*, 2014, **136**, 7555.
67. B. Huang, Z. Y. Li, Z. R. Liu, G. Zhou, S. G. Hao, J. Wu, B. L. Gu and W. H. Duan, *J. Phys. Chem. C*, 2008, **112**, 13442-13446.
68. A. Saffarzadeh, *J. Appl. Phys.*, 2010, **107**, 114309.
69. T. H. Osborn and A. A. Farajian, *Nano Res.*, 2014, **7**, 945-952.
70. J. Taylor, H. Guo, and J. Wang, *Phys. Rev. B,* 2001*,* **63**, 245407.
71. M. Brandbyge, J.-L. Mozos, P. Ordej_on, J. Taylor, and K. Stokbro, *Phys. Rev. B*, 2002, **65**, 165401.

72 Atomistix Toolkit version 2015.0; QuantumWise, Copenhagen, Denmark; see http://www.quantumwise.com/
73 S. Grimme, *J. Comp. Chem*., 2006, **27**, 1787.
74 S. Grimme, C. Mück-Lichtenfeld, and J. Antony, *J. Phys. Chem, C*, 2007, **111**, 11199-11207.
75 F. Boys and F. Bernardi, *Mol. Phys*., 1970, **19**, 553.
76 M.E.Dávila, A.Marele, P. De Padova, I. Montero, F. Hennies, A. Pietzsch, M. N. Shariati, J. M. Gómez-Rodríguez and G. LeLay, *Nanotechnology*, 2012, **23**, 385703.
77 M. Y. Han, B. Özyilmaz, Y. B. Zhang and P. Kim, *Phys. Rev. Lett.*, 2007, **98**, 206805.
78 Z. Chen, Y.-M. Lin, M. J. Rooks and P. Avouris, *Physica E*, 2007, **40**, 228–232.
79 L. B. Luo, X. B. Yang, F. X. Liang, J. S. Jie, C. Y. Wu, L. Wang, Y. Q. Yu and Z. F. Zhu, *J. Phys. Chem. C*, 2011, **115**, 24293-24299
80 J. S. Park, B. Ryu, C. Y. Moon and K. J. Chang, *Nano Lett*.,2010, **10**, 116-121.
81 L. B. Luo, X. B. Yang, F. X. Liang, H. Xu, Y. Zhao, X. Xie, W. F. Zhang and S. T. Lee, *J. Phys. Chem. C*, 2011, **115**, 18453-18458.
82 K.-H. Hong, J. Kim, J. H. Lee, J. Shin and U. I. Chung, *Nano Lett.*, 2010, **10**, 1671-1676.